# Dynamic interactions between pancake vortex stacks and Josephson vortices in $Bi_2Sr_2CaCu_2O_8$ single crystals: relaxation and ratchets

G.K.Perkins, A.D.Caplin and L.F.Cohen

*Center for Electronic Materials and Devices, Blackett Laboratory, Imperial College, London SW7 2BZ, UK*

*We present a detailed study of the dynamic interactions between Josephson vortices and stacks of pancake vortices in a $Bi_2Sr_2CaCu_2O_8$ single crystal, obtained by measuring the effect of applied in-plane magnetic field pulses on the c-axis magnetisation. The predominant interaction is to relax the system towards equilibrium. However, using a highly sensitive AC technique we are able to measure also the forces acting to drive the system away from equilibrium, consistent with the existence of dragging interactions between the Josephson and pancake systems. Such forces were discussed recently as the basis of possible flux ratchet devices.*

Due to its highly anisotropic layered structure, the high temperature superconducting cuprate $Bi_2Sr_2CaCu_2O_8$ exhibits a wide array of vortex phenomenology, particularly for the case of an applied magnetic field that is tilted away from the sample c-axis. Above a critical angle the conventional tilted pancake vortex (PV) stacks are replaced by a composite system of two sub-lattices, comprising an array of Josephson vortices (JV) running parallel to the crystallographic ab planes, while the stacks within the PV lattice remain close to the c-axis.[1,2,3] The interactions between the two sub-lattices lead to further vortex phases within the composite state[4] which persist to temperatures above 77K.[5]

It has been observed too that if JV are moved repeatedly in and out of the sample, the *c*-axis magnetic moment relaxes rapidly to its equilibrium state.[5] Most interestingly, this interaction was recently suggested as a possible mechanism for rectification of flux motion,[6] which could form the basis superconducting electronic components such as diodes or flux pumps. Similar 'vortex ratchets' have been considered before,[7,8] but their practical realisation requires the controlled fabrication of asymmetric pinning potentials, which is difficult to accomplish in practice. However, the recent proposal uses the *intrinsic* interactions between the JV and PV to provide the rectification. The central argument is that the JV exert a dragging force on the PV as they pass through the sub-lattice and that this force is dependent on the JV velocity. Therefore driving JV with a time-asymmetric waveform should impose net bias in one particular direction, causing rectification. Although there have been several theoretical studies of the JV-PV interaction, very little experimental work has been done, particularly with regards to the dynamics, so the existence of JV-PV dragging interactions has not previously been confirmed.

Here we present an experimental investigation of the details of the dynamic interactions between the two vortex systems by studying the effect of field pulses directed along the ab-planes (driving the JV sub-lattice) on the c-axis (pancake) magnetisation of a $Bi_2Sr_2CaCu_2O_8$ single crystal. We discuss first the relaxation effects, which dominate when the sample is far from equilibrium, and then show evidence of dragging forces with the system close to equilibrium.

A $Bi_2Sr_2CaCu_2O_8$ crystal (~500x500x20μm) was mounted onto a sapphire block and immersed in liquid nitrogen, hence all the measurements reported here correspond to a sample temperature of 77K. The c-axis magnetisation was probed using a scanning Hall sensor (200μm InSb Hall bar of sensitivity ~1mGs/√Hz) where the magnetic induction $B_{centre}$ was recorded with the sensor positioned 150μm above the centre of the sample surface (so that the sensor effectively averages the induction over an area of a few hundred microns in the centre of the sample). The external induction $B_{external}$ was recorded with the sensor positioned 500μm away from the sample edge, so that the sample self-induction $B_{self} = B_{centre} - B_{externa}$ is proportional to the c-axis component of the sample magnetisation $M_c$ (the scaling factor can be estimated from the sample-sensor geometry with an uncertainty of ~30%).

Two sets of split-pair coils generate in-plane, and c-axis, applied magnetic fields ($H_{ab}$ and $H_c$ respectively). The former is used to provide triangular $H_{ab}$ pulses up to an amplitude of 10Oe with variable rise and fall times in the range 10μs to 10ms while the latter provides a DC field in the range –30Oe<$H_c$<30Oe. It is crucial to minimise and quantify any misalignment of the ab field. Therefore we compared the virgin slopes of the magnetisation curves for $M_c$ independently as a function of $H_c$ and $H_{ab}$ (any uncertainty in the relationship between $B_{self}$ and $M$ is the same in both cases and hence cancels out in the estimation). In the latter case, the only signal arising is from any misalignment. We found the ratio of the slopes to be about 150:1, corresponding to a field misalignment of less than 0.5º. The effect of this c-axis contamination can be gauged by observing the effect of c-axis pulses of a similar amplitude generated by the c-axis coil. In general they have no observable effect on the magnetisation (as expected from the critical state model when the c-axis pulse amplitude is very small compared to the c-axis penetration field). Hence we can be confident that the effects reported here are solely a result of the ab-pulses, and not artefacts of c-axis contamination. Also, we confirmed, by varying the pulse rate by an order of magnitude, that there were no significant sample heating effects.

The predominant effect of the in-plane pulses is to relax the system towards equilibrium. Fig.1 shows the magnetisation loops $B_{self}$ vs. $H_c$ in the absence of any $H_{ab}$ pulses and with 1000 $H_{ab}$ pulses applied before each data reading ($B_{self-relaxed}$). While the former clearly shows hysteresis, the latter shows none within the experimental resolution, indicating that the system is close to equilibrium. The initial legs (corresponding to the Meissner regime) of both $B_{self}$ and $B_{self-relaxed}$ are collinear as expected, but because the sensor is at a finite height above the sample, the initial slope $dB/dH$ is reduced from the fully-diamagnetic -1 to -0.7. The solid line is $B_{self}$ - $B_{self-relaxed}$ (i.e. the irreversible component) for both the upward and downward legs and shows a clear asymmetry (the downward leg being much greater in magnitude than the upward leg). Such behaviour is consistent with surface or geometric barriers acting to inhibit flux exit from the sample.

We conducted a detailed investigation of the relaxation process by measuring $B_{self}$ after successive $H_{ab}$ pulses. The sample is prepared in the critical state by cycling the applied magnetic field $H_c$ to the target field. A waiting period of 1 minute is then allowed for thermally-activated creep to fall to a negligible rate. In all cases the residual thermal creep that would otherwise take place during the period of a measurement is less than 1%, as checked directly in a separate set of thermal creep experiments. After the waiting period, triangular ab-pulses of duration 2ms (1ms each for the rise and fall times) were applied at a rate of 10 per second, and the signal $B_{centre}$ recorded after each pulse. Typically 500 pulses are applied in this manner, then a final burst of 1000 pulses applied at a rate of 100 per second. This is sufficient to completely relax the system to the reversible, equilibrium, state as previously demonstrated (fig.1). A final measurement of $B_{centre}$ is then made, which is subtracted from the whole dataset leaving just the non-equilibrium (irreversible) component of the sample self-field $B_{irr}$. Examples of the relaxation curves are shown in fig.1 (inset), which are quasi-exponential in nature. Induced relaxation of the critical state by a transverse AC field was reported earlier.[9] However, a theoretical discussion has only recently been proposed by Brandt et al.[10] Using a 3D model of line-like vortices (as opposed to PV stacks), the relaxation was described in terms of a 'tilt' mediated vortex motion. This results in a predicted exponential decay, defining a relaxation rate $R = d\ln B_{irr}/dP$, similar to our observations, but it is unlikely that this model is appropriate for a strongly-layered system such as $Bi_2Sr_2CaCu_2O_8$, especially in the light of the recent evidence for a composite vortex lattice structure. For a slab of aspect ratio $\eta$, Brandt's model provides a semi-quantitative prediction $R = 8A/(\eta H_p)$ (we have converted from units of frequency to pulse$^{-1}$), where the ab pulse amplitude $A$ is large compared the critical state penetration field $H_p$. In our experiments $A$ is typically 10Oe and $H_p$ is less than 1Oe so this limit is appropriate. However, this predicts $R\sim1$, some three orders of magnitude greater than our measured values.

Clearly an alternative interpretation of the data is needed, and it should be one taking into account the composite JV-PV lattices and their dynamic interactions. Within such a framework the exponential decay of the irreversible magnetisation corresponds to the amount of PV flux entering or exiting at each pulse being proportional to that magnetisation, i.e. to the number or density of PVs. A simple picture is that the driven JV induce a flux-flow like state in the PV ensemble, and a mechanism by which this could occur is depinning of the PV by the formation of kinks at the PV-JV intersections.[4]

A comprehensive theoretical treatment must account for the complex collective interactions between the two systems of vortices. In particular the observed field dependence and pulse-amplitude dependence of $R$ (fig. 2) point towards such collective effects. Here, $R$ is observed to increase with increasing field ($H_c$) on the upward leg of the $M$-$H$ loop while on the downward leg $R$ decreases with increasing field. This asymmetry highlights the influence of surface/geometric barriers on the downward leg, which suppress the depinning process (it is more difficult for the PVs to relax out of the sample), while the relatively strong field dependence on the upward leg is a signature of collective relaxation processes. Furthermore, the pulse amplitude (proportional to the JV density) dependence (fig. 2 inset (a)) appears to be quadratic, faster than the expected linear behaviour that would result from single-vortex-like depinning (simply proportional to the number of JV-PV intersection events). Finally the pulse duration dependence shown in fig.2 (inset b) addresses the JV velocity dependence of the JV-PV interaction, and shows a logarithmic increase in $R$ for increasing pulse duration, indicating a stronger interaction for slower-moving JV. This logarithmic behaviour may be related the collective processes observed in standard thermal creep measurements, which are also observed to be quasi logarithmic.

The results described above provide evidence of interactions between the PV and JV, and reflect processes that relax the system towards equilibrium (i.e. a depinning interaction). However, they do not show any *direct* evidence of dragging interactions between the two sets of vortices. The depinning effects may well be useful, for example, to remove unwanted fluxons from various fluxonic devices, but for flux ratchet devices it is necessary to be able to drive the system away from equilibrium, possibly by utilising JV-PV dragging interactions. To address this issue, we have developed a highly-sensitive AC technique to probe the sample response close to equilibrium. In these experiments we apply a continuous sinusoidal AC field at frequency $f$ in the ab direction. The dragging forces between JV and PV depend on the direction of motion of the JV, but not on their polarity, and so have a frequency $2f$ and lag in phase of $90^O$ (fig.3). Hence $B_2''$, the quadrature component at $2f$, is a signature of dragging interactions. Furthermore, because the signal is modulated the sensitivity is greatly improved and detection at the second harmonic effectively rejects much of the background AC coupling in the system electronics. Consequently our minimum detectable signal with this method is ~10μGs (2 orders of magnitude more sensitive than in the relaxation protocol).

The data (fig. 4) show a clear signature with sharp peak in $B_2''$ at around $H_c = \pm 2$Oe (there is a slight offset 0.5Oe in $H_c$ due to the background magnetic field) indicating a maximum in the PV displacement at these fields. We have measured the phase angle to be ~-70$^O$ showing that the PV response is roughly synchronous with the function abs(f) as expected. This is consistent with a JV-induced compression of the PV system resulting from a dragging interaction between the two sets of vortices. Furthermore, the subsequent suppression of $B_2''$ with increasing field may be due to the increasing elastic modulii of the PV lattice.

Our interpretation of these data is strengthened by consideration of the following key points: Firstly, no such effect is observed if the Hall sensor positioned 500μm away from the sample, proving that the signal $B_2''$ is generated by the sample itself and is not some electronic artefact. Secondly we can be sure that the measured signal reflects the PV response and not the JV response because the data are antisymmetric around zero field, i.e. for anti-PV (negative applied c-axis field) the signal is of the opposite sign as for normal PV (positive fields). Finally, to check that the PV response results from the driven JV vortices and not some artefact due to the small c-axis component present in the ab pulses we repeated the experiment applying just a c-axis AC field of amplitude 0.15Oe (rms). This amplitude corresponds to three times the measured c-axis contamination of the ab pulses as described earlier. Although there is a detectable signal with a similar background trend (which could explain the structure observed in $B_2''$ for fields above the peak field), it is without the sharp peaks seen with the ab pulses. We can therefore be confident that the observed peaks represent the PV response to the driven JV system, which is of the correct frequency and phase for the hypothesised dragging forces between the JV and PV. Although the signals are very small, of order 1mGs and so about 3 orders of magnitude smaller than the applied c-axis field, they correspond to a strain in the PV lattice of ~0.1%, or displacements at the sample periphery of ~1μm (large enough for the operation of nano scaled fluxonic devices). For comparison, the spacing between PV at the applied field of the peaks is about 5μm.

In conclusion we have investigated the dynamic interactions between pancake vortex stacks and driven Josephson vortices in a $Bi_2Sr_2CaCu_2O_8$ single crystal by analysing the relaxation of the c-axis magnetisation induced by successive field pulses directed along the ab planes. Far from equilibrium the predominant interaction is that of depinning leading to exponential relaxation. The relaxation rate depends quadratically on pulse amplitude and logarithmically on the pulse duration. On the other hand when the system is close to equilibrium, we have observed evidence for dragging interactions between the two sets of vortices. These results indicate that an in-plane AC field of amplitude 10Oe and frequency 1kHz induces micron scale displacements in the PV lattice. Such an effect could form the basis of recently proposed superconducting electronic devices based on vortex ratchets.

### Figure Captions

Fig.1 ***M-H*** loops measured using a scanning Hall magnetometer. Open squares, the sample self induction -obtained by subtracting the induction measured with the Hall sensor positioned 500μm away from the sample from that measured with the sensor positioned 150μm above the centre of the sample. Filled squares, the reversible part of the sample induction $B_{self-relaxed}$ measured in a similar fashion but after each c-axis field step 1000 ab pulses are applied before the induction $B_{self}$ is measured

(resulting in a reversible *M-H* curve). Solid line, the irreversible part of the magnetisation $B_{self}$-$B_{self-relaxed}$ showing significantly different behaviour on the upward and downward legs indicating the presence of surface or geometric barriers. (inset) examples of the relaxation curves showing sample magnetisation $B_{self}$ as a function of the number of applied ab field pulses at applied fields of 0Oe and 1Oe (for both the upward and down legs of the *M(H)* cycle). The relaxation is observed to be quasi exponential.

Fig.2. The relaxation rate *R* versus $H_c$ for the upward (closed squares) and downward (closed circles) legs of the *M-H* loop, showing a rapid increasing field dependence for the upward leg and a slower decreasing dependence on the downward leg. Inset (a) The dependence of *R* on pulse amplitude (at $H_c$=0), which follows a quadratic dependence. Inset (b) The dependence of *R* on pulse duration (at $H_c$=0), showing a logarithmic increase in *R* with longer pulse duration.

Fig. 3. JV are driven through the sample by an ac field applied in the ab direction at frequency *f*. Because dragging forces do not depend on the JV polarity, only on the direction of motion it is the function abs(*f*) that represents the driving force on the PV. This is similar to an AC drive at 2*f* with a phase lag of $90^O$.

Fig 4 Out of phase component at the 2nd harmonic of the c-axis induction when a 1kHz, 7Oe rms AC field is applied in the ab direction (closed squares), and with a 1kHz, 0.15Oe rms AC field applied in the c-direction (open squares). A pronounced peak is observed for the ab modulation only, consistent with dragging interactions between the JV and PV. When the Hall sensor is moved away from the sample by 500μm (indicated 'off sample'), the signal is small and constant. The sign of the peak is negative for positive c-axis field, showing the phase of the signal to be negative, consistent with dragging interaction between the PV and JV.

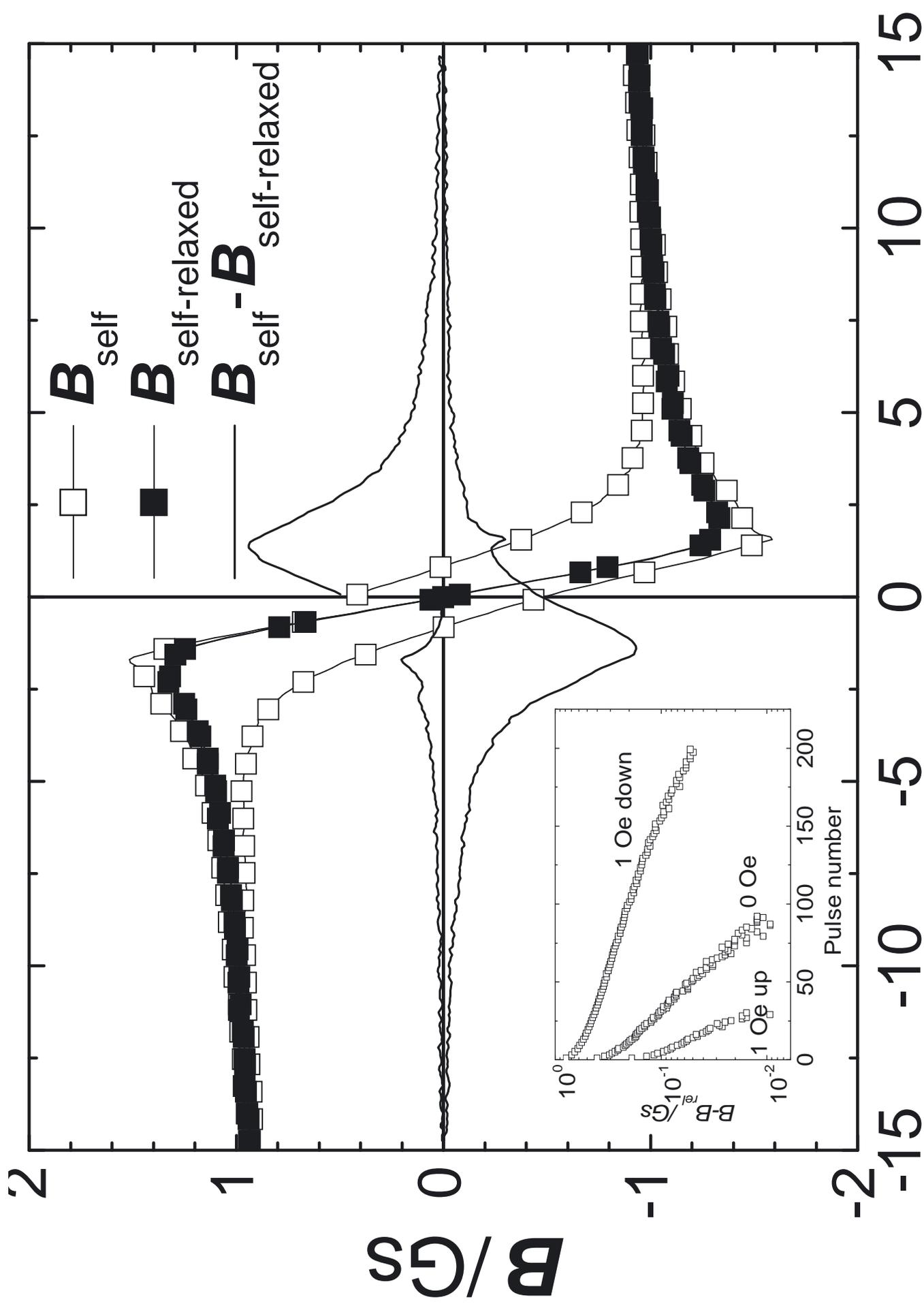

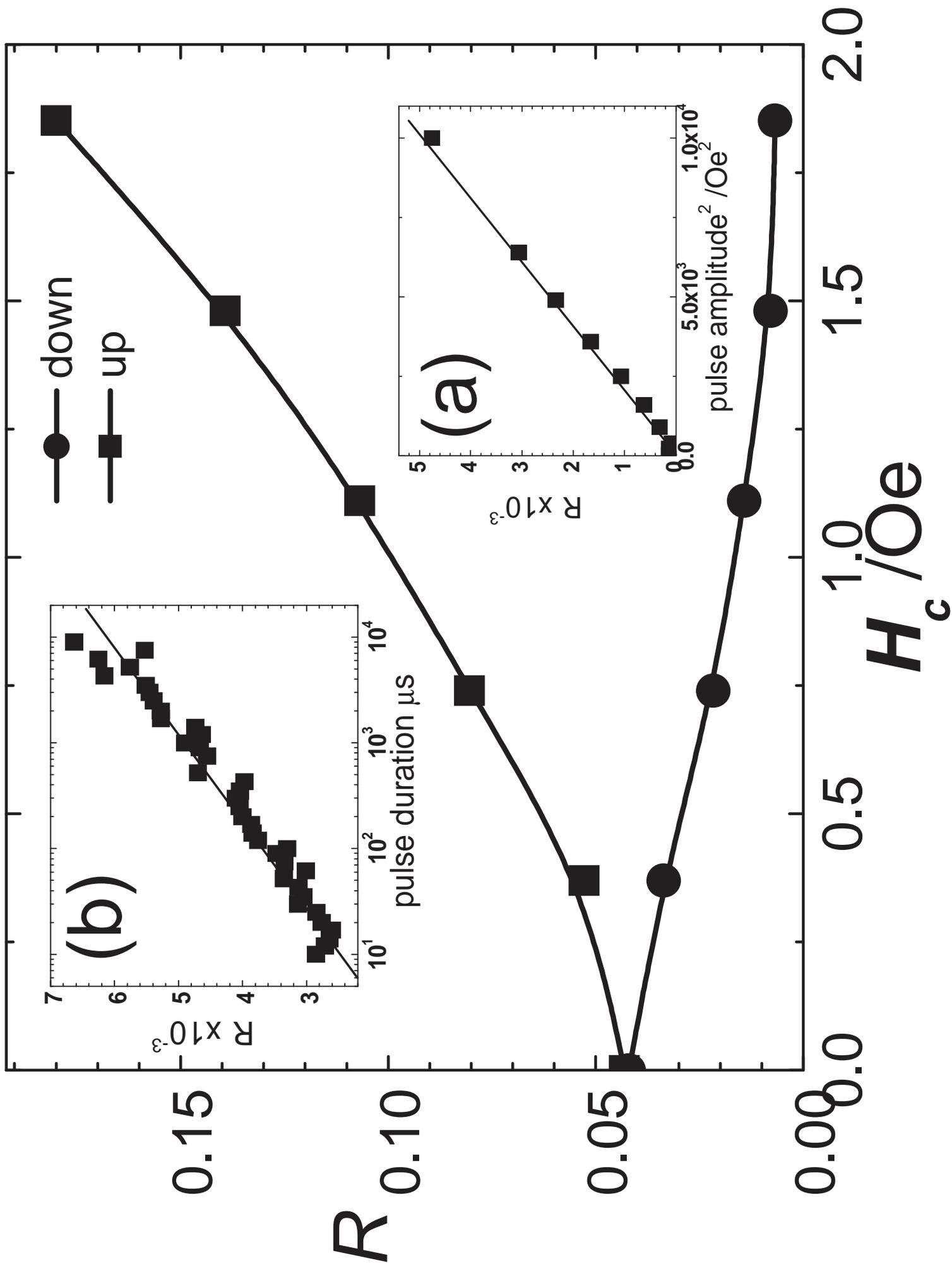

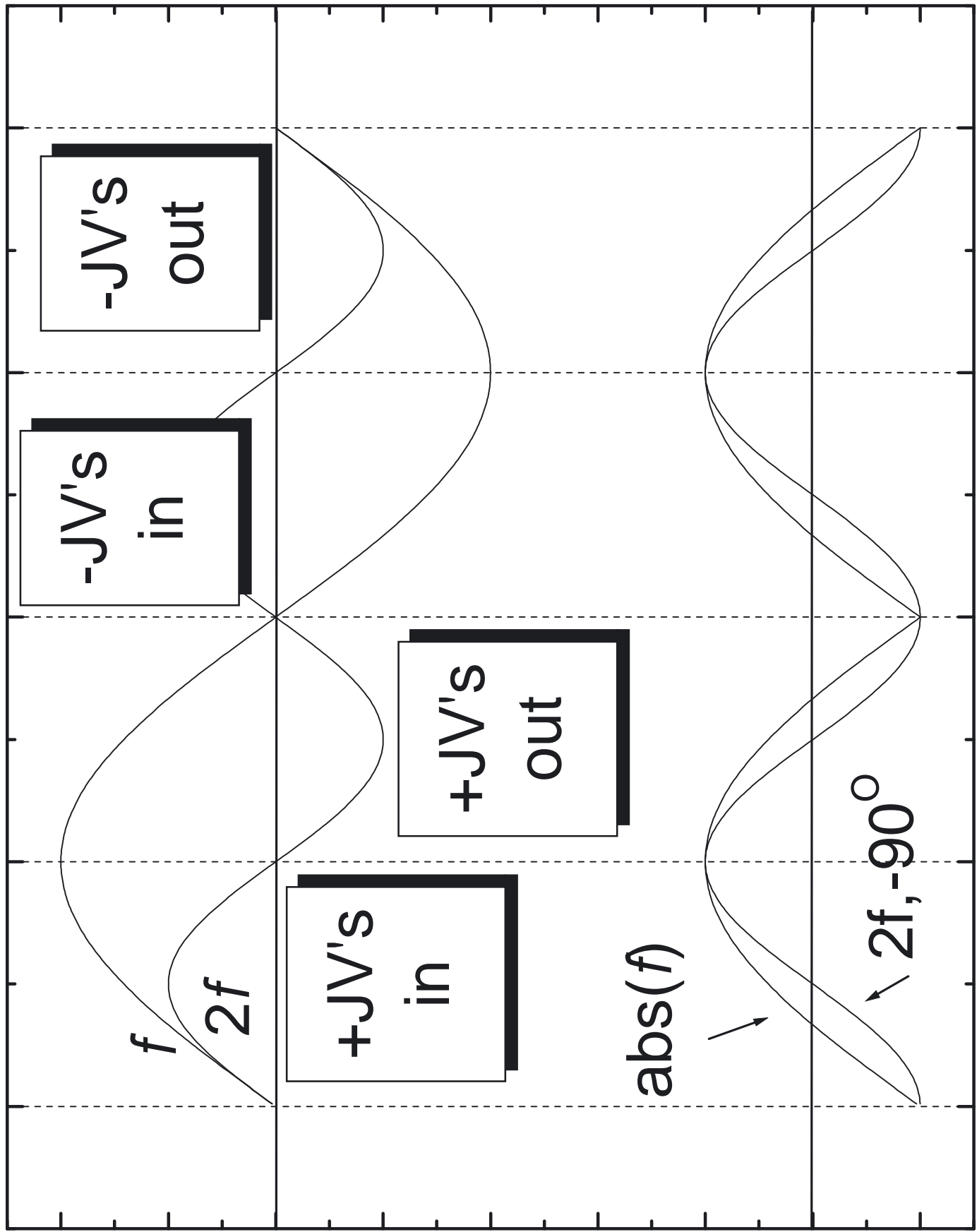

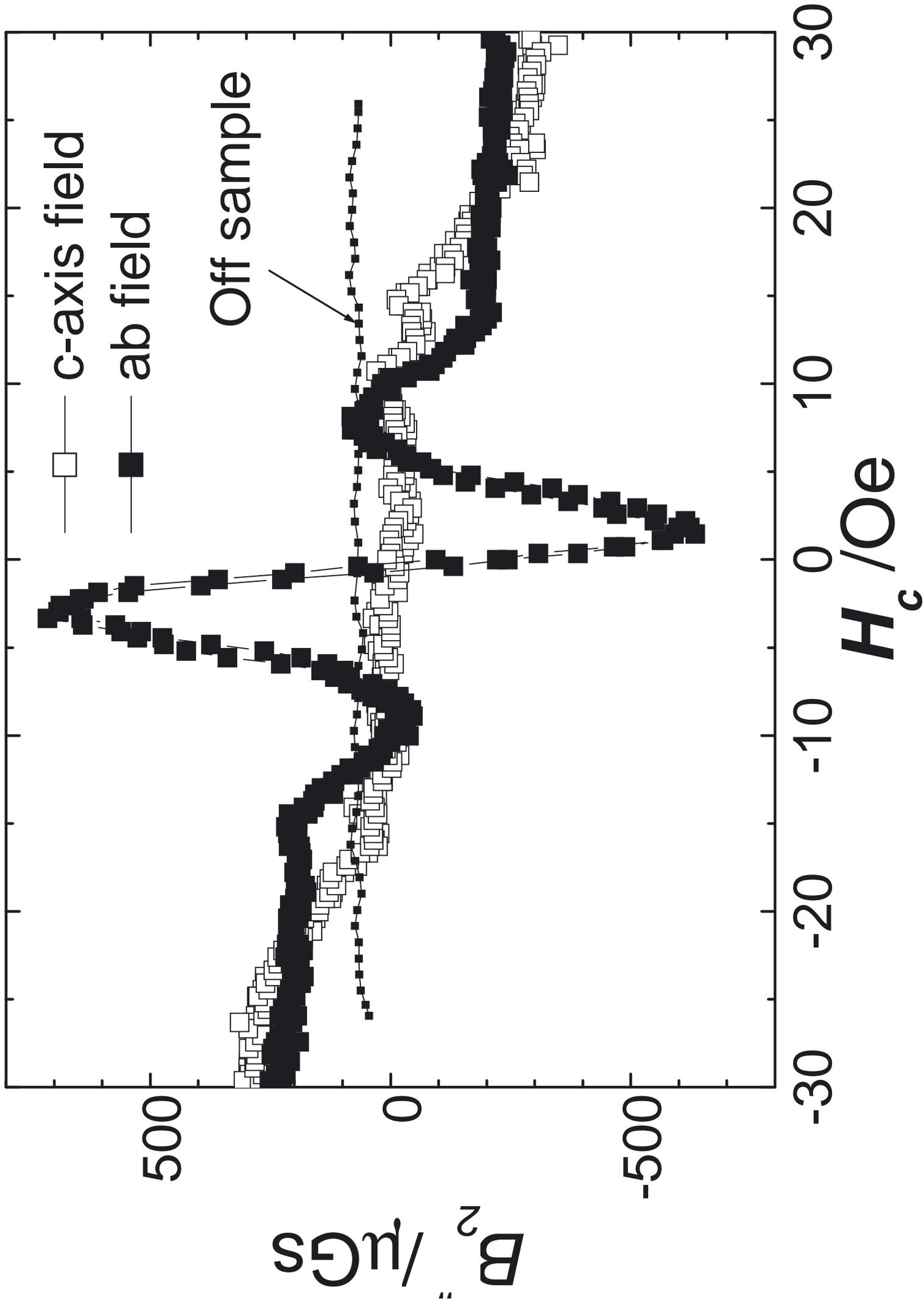